# Global-Scale Resource Survey and Performance Monitoring of Public OGC Web Map Services

**Zhipeng Gui** [1,*], **Jun Cao** [2,3], **Xiaojing Liu** [2,3], **Xiaoqiang Cheng** [2,3] **and Huayi Wu** [2,3,*]

[1] School of Remote Sensing and Information Engineering, Wuhan University, 129 Luoyu Road, Wuhan 430079, China
[2] State Key Laboratory of Information Engineering in Surveying, Mapping and Remote Sensing, Wuhan University, 129 Luoyu Road, Wuhan 430079, China; caojun1212@whu.edu.cn (J.C.); liuxiaojing@whu.edu.cn (X.L.); carto@whu.edu.cn (X.C.)
[3] Collaborative Innovation Center of Geospatial Technology, Wuhan University, 129 Luoyu Road, Wuhan 430079, China
* Correspondence: zhipeng.gui@whu.edu.cn (Z.G.); wuhuayi@whu.edu.cn (H.W.);
  Tel.: +86-027-6877-7167 (Z.G.); +86-027-6877-8311 (H.W.)

**Abstract:** One of the most widely-implemented service standards provided by the Open Geospatial Consortium (OGC) to the user community is the Web Map Service (WMS). WMS is widely employed globally, but there is limited knowledge of the global distribution, adoption status or the service quality of these online WMS resources. To fill this void, we investigated global WMSs resources and performed distributed performance monitoring of these services. This paper explicates a distributed monitoring framework that was used to monitor 46,296 WMSs continuously for over one year and a crawling method to discover these WMSs. We analyzed server locations, provider types, themes, the spatiotemporal coverage of map layers and the service versions for 41,703 valid WMSs. Furthermore, we appraised the stability and performance of basic operations for 1210 selected WMSs (*i.e.*, *GetCapabilities* and *GetMap*). We discuss the major reasons for request errors and performance issues, as well as the relationship between service response times and the spatiotemporal distribution of client monitoring sites. This paper will help service providers, end users and developers of standards to grasp the status of global WMS resources, as well as to understand the adoption status of OGC standards. The conclusions drawn in this paper can benefit geospatial resource discovery, service performance evaluation and guide service performance improvements.

## 1. Introduction

Web technologies, new standards and commercial applications are rapidly changing the nature and extent of available online geospatial resources. Thus, it is imperative to investigate whether Web Map Service (WMS) remains the best solution to share and interoperate maps over the Internet or if a new standard and direction is needed. To understand the WMS adoption situation and appraise the WMS standard, a global-scale survey to investigate the distribution, usage and quality of public WMSs (*i.e.*, the web map services that are available over the Internet for public use) is highly desirable. WMS is an Open Geospatial Consortium (OGC) standard protocol initially proposed in 2000 for serving geo-referenced maps over the Internet [1]. The latest version WMS 1.3.0 was published in 2006 [2] and is also available as ISO 19128 (*i.e.*, ISO 19128:2005 Geographic information-Web map server interface) [3] issued by the International Organization for Standardization (ISO). WMS has become the most widely-used OGC data portrayal service, recognized globally and supported by both mainstream commercial geospatial tools and open-source software. Currently, a huge number of WMSs are available over the Internet for public use providing abundant map resources, but varying in map content and quality. In this context, finding an appropriate WMS becomes a challenging problem [4].

Therefore, for more effective and efficient utilization of these invaluable online geospatial resources, the following questions need to be asked: What is the adoption situation of WMS; who are the primary contributors; and where are the providers located? Are there any patterns regarding the different attributes of WMS resources? For example, what are the most popular themes, specification versions and map projections, spatial and time coverages of the map layers? How is the quality of global WMSs, in terms of accessibility, successability and performance? Are there any patterns that can provide hints for service selection and server-side improvement?

In this paper, we carried out a global-scale resource investigation and executed distributed performance monitoring on a set of crawled public WMSs. The potential contributions of this research can be drawn from the following perspectives:

(1) Geospatial resource discovery: the investigation of WMS resources and their metadata help us to grasp the server locations, provider types and content distribution of the global geospatial resources. Consequently, this knowledge will benefit service discovery by bringing to light on-demand WMSs that have those properties expected by service consumers.
(2) Service performance evaluation: by developing a distributed monitoring framework, spatiotemporal heterogeneity and individual service-level performance differences can be analyzed over space and time. This will guide performance-aware service selection for time-critical applications, as well as steer performance improvements from the service provider perspective.
(3) Evolution of service standard: this investigation can help researchers and standard makers from both industry and academia review the development and adoption of open service standards for a geospatial data portrayal. By linking cutting-edge web visualization technologies in relation to the prevailing interoperation modes (e.g., crowdsourcing [5,6] and collaboration [7,8]), we can rethink the appropriateness of the WMS standard, thus inspiring us to conceive new directions for advancement.

This article is organized as follows. Section 2 reviews relevant research. Section 3 introduces our monitoring method and data collection. Section 4 details the discoveries. Section 5 concludes with results and discusses future research.

**2. Related Work**

*2.1. Development of Online Mapping Technologies and WMS*

With the development of web technologies, interactive online mapping has made great progress over the past several decades. Successively emerging commercial map services and crowdsourcing projects, like MapQuest, Open Street Map and Google Maps that implemented various technologies and open standards are significantly changing the application of online mapping [9]. Online mapping relies on rendering strategies, data models and markup languages; nowadays, both server-side rendering and client-side rendering play important roles in online mapping. Server-side rendering generates maps (typically raster images with vector objects and annotations overlaying) on map servers and delivers them to clients. Since it relies on the server-side functionality and computing powers, the access and presentation of maps became much easier without any advanced processing on the client side. Server-side rendering, however, results in poor interactivity. Intensive concurrent map requests introduce frequent data conversion and transfer issues and may lead to server-side overload and low responsivity on the client side [10]. Newly-developed data cube technologies and algorithms enable interactive exploration of large multidimensional spatiotemporal datasets (billions of entries) with very low latencies [11], e.g., nanocubes. In comparison to server-side rendering, client-side rendering enables data rendering, animation and enriched interaction features directly in web browsers. Datasets, instead of rendered images, are retrieved for the client. Data models based on Extensible Markup Language (XML) and JavaScript Object Notation (JSON) prevail, such as Scalable Vector Graphics (SVG), Geography Markup Language (GML) and GeoJSON. Conventional plugin-based Rich Internet Application (RIA) technologies (e.g., Adobe Flash, Microsoft Silverlight and Oracle-Sun JavaFX) have wide applications [12], but also disadvantages [13,14], such as extra installation, security and compatibility concerns. With the evolution and standardization of web technologies, native HyperText Markup Language (HTML) 5.0 and JavaScript packages have expanded to include powerful visualization and rendering functionalities. Other open standards, such as HTML Canvas and WebGL [15], also show great potential for data rendering. These standards and technologies have alleviated the dependence on plugins and make data rendering more interactive and flexible.

The WMS standard relies on a server-side rendering mode, thus the maps are generated by a map server using geospatial data from geospatial databases or other data sources. WMS defines a set of standardized operations (*i.e.*, interfaces) to facilitate map requests. *GetCapabilities* accesses the metadata of the service and map layers; *GetMap* generates a map with well-defined geographic and dimensional parameters; and *GetFeatureInfo* retrieves extra properties for particular features shown on a map [2]. These operations can be integrated with other geospatial tools; or invoked using a standard web browser by submitting HyperText Transfer Protocol (HTTP) requests. It is easy to request images with changing parameters on-demand for clients.

Currently, WMS is being challenged by competitors, such as the Tile Map Service Specification (TMS) [16] and ESRI RESTful services. The stability and the performance of a map server deteriorate rapidly when high concurrency and frequent interaction occurs. To tackle this problem, various approaches have been developed and are widely-used, e.g., message queue, indexing, auto-scaling and dynamic load balancing technologies. Beside these methods, map tile caching mechanisms provide an alternative approach to fetch and cache pre-rendered map tiles efficiently, according to specified geographical extent and scales. TMS is one of the earliest standards for tiling maps; gaining a great deal of support from many open source communities. As Representational State Transfer (REST) advances as the mainstream architectural style for web services, ESRI has proffered RESTful Service Application Programming Interfaces (APIs) starting in 2010. With the vast deployments of the ArcGIS Service architecture, a growing number of ESRI RESTful services are available over the Internet. To meet these challenges and ever-changing demands, OGC constantly improves its standards. Inspired by TMS, OGC published the Web Map Tile Service (WMTS) standard in 2009 to develop scalable, high performance services for web-based distribution of maps [17]. WMTS provides a complementary approach to WMS for tiling maps. Moreover, in 2011 OGC formed a working group to explore the implementation of geospatial services through RESTful approaches [18].

OGC standards have their own advantages when compared to the industrial web service standards advocated by the World Wide Web Consortium (W3C). OGC service standards provide abundant metadata about the provider and the geographic data through *GetCapabilities* and the capability is continuously enhanced. In contrast, Web Service Description Language (WSDL) only focuses on syntactic service APIs and message interaction. For example, WMS has supported the description and access to time series data through dimension parameters since Version 1.1.0. Data animation functions can be implemented on the client side to visualize the dynamics of natural phenomena or socioeconomic processes conveniently [19]. Contemporary commercial and open-source map APIs (e.g., Google Maps, Bing Maps and OpenLayers) provide capabilities to integrate server-side and client-side mapping in applications. An integrated solution combining WMSs with vector data overlays provides a productive choice for advanced users [15]. In summary, WMS plays a crucial role in online mapping and is widely supported by both commercial (e.g., ArcGIS products [20], Autodesk's Map 3D [21] and Civil 3D products [22]) and open-source (e.g., OpenLayers [23], GRASS GIS [24], QGIS [25]) software providers and various applications.

Meanwhile, there are urgent and evolving demands for the standard to be more interactive, analytic and collaborative [26].

*2.2. Online Geospatial Web Service Survey*

Investigating global online geospatial web services is essential for geospatial resources discovery and a better understanding of the status of online resources [27,28]. Spatial Data Infrastructures (SDIs) have been widely used in Earth science domain to facilitate geospatial resources discovery and sharing [29]. Catalogues and portals, such as Global Earth Observation System of Systems (GEOSS) Clearinghouse [30] and data.gov [31], maintain millions of metadata entries for geospatial resources, allowing users to specify query constraints by using indexed metadata property fields. An SDI-level geospatial resource survey would provide invaluable information for both geospatial resource producers and consumers, even for policy makers. However, most of the SDIs only provide coarse-grained statistics, such as available resource types and regional-level resource distributions. A detailed publicly-available, resource survey of global geospatial web services has never been executed or, alternatively, is not available to end-users. Furthermore, registry records may be stale and incomplete in SDIs because the metadata maintenance depends on service owners. Lopez-Pellicer *et al.* [32,33] analyzed the discovery capability of common search engines and SDIs toward OGC services. Common search engines, like Google, Yahoo and Bing, can only index and recall half of the OGC services at most, and SDIs have limited resource coverage. Therefore, an active resource investigation is required.

There are many available online WMS lists provided by third parties. Refractions Research [34] collected 615 WMSs by using Google Web APIs and extracted the basic metadata fields (e.g., name, title, layer bounding box). Skylab Mobilesystems Ltd. [35] also offers a frequently-updated list of unrestricted accessible WMSs at the global extent, but only the number of layers was counted; furthermore, the number of WMSs was limited (994 WMSs). To understand the provenance of online OGC web services, the geo-distribution of service providers [36] and the service deployment situation (e.g., the number of services deployed on each server and the number of dataset provided by each service) were studied [32,37]. This research revealed the imbalances in geospatial resources in terms of service location and provider. By analyzing the service types and version proportion of online OGC services in Europe, Lopez-Pellicer *et al.* [32] found that WMS is the most popular OGC service online since it is easy to deploy and use. However, the maintenance and updates of these online WMSs were inefficient. Li *et al.* [37] investigated the web diffusion of WMS by developing an active crawler, determining that the total number of WMSs continuously increased, while at the same time, some WMSs became invalid (*i.e.*, the access URL become invalid or the *GetCapabilities* operation cannot properly response constantly). Thus, the maintenance and stability of online WMSs are big issues.

So far, there has been little research conducted from the perspective of the map contents provided by WMSs. For the automatic detection of orthoimage layers offered by WMSs, Florczyk *et al.* [38] proposed a heuristics method that combines both capabilities of document description-based analysis and content-based computation together. This research has been integrated with the Virtual Spain project to benefit Digital Elevation Model (DEM) and realistic 3D view generation. Current research provides pioneering, but limited work. The major drawbacks are as follows: (1) none have conducted a global-scale resource survey, and the number of investigated WMSs discussed in the literature was small; (2) the analysis of the service content was limited. Only a few metadata properties were studied. Rarely does the existing research explain and discuss discoveries in relation to policy and technical issues. So far, we still only have limited knowledge of the global distribution of WMS servers, provider types or content (e.g., primary map subjects, data collection times and spatial coverage). The adoption and usage status of WMSs are also unclear, such as the most frequently-used service version, the updating status of services, map content or the widely-used Coordinate Reference System (CRS). Therefore, a global-scale investigation to grasp the resource distribution and adoption status of OGC standards is urgently needed.

*2.3. Quality of Service Monitoring*

The *GetCapabilities*, *GetMap* and *GetFeatureInfo* operations accomplish the functional requirements supposed to be implemented by a service during interoperation. In contrast, non-functional requirements, so-called Quality of Service (QoS) attributes, such as reliability, maintainability and performance, measure the overall properties of web services [39]. The Infrastructure for Spatial Information in the European Community (INSPIRE) established QoS requirements for spatial dataset viewing services. INSPIRE insists that QoS criteria, such as performance, capacity and availability, shall be ensured for regulatory requirements [40,41]. To evaluate the QoS offered by service providers, quality monitoring becomes imperative and urgent. Among these QoS criteria, availability and performance especially are expected by the user, since they explicitly impact the user experience [42].

To acquire quality data, various monitoring methods have been proposed. General test tools, such as Apache JMeter and LoadRunner, provide powerful capabilities for conventional performance and load tests. Utilizing these tools, experiments have been conducted to analyze the key performance factors of OGC web services [43–45]. However, general tools cannot parse service-specific data packages. As a result, the metadata of geospatial web services cannot be extracted, and advanced monitoring cannot be conducted automatically [42]. To address this issue, domain-oriented monitoring infrastructures were proposed. The Federal Geographic Data Committee (FGDC) established the Service Status Checker (SSC) to verify and grade various types of geospatial web services [46]. MapMatters is a monitoring platform providing a web portal to visualize the quality of WMSs, exclusively [42]. To facilitate geospatial resources discovery, Li *et al.* [47] and Gui *et al.* [48] designed one-stop discovery portal prototypes to integrate service performance monitoring and visualization functions. In order to alleviate the loading burden on the monitored servers, Wu *et al.* [49] proposed a flexible framework to adjust the monitoring time interval dynamically according to the recent performance of selected services. Currently, most of the existing monitoring frameworks employ a single monitoring site mode and sparse monitoring time intervals. However, web application performance varies in space and time, impacted by many factors [50]. Therefore, the performance data collected from one geo-location at one fixed time point cannot describe the performance in another geo-location or at another time. Biased monitoring data may mislead quality evaluation and service selection [50,51].

Although geospatial web service monitoring and analysis have yielded abundant outcomes, the following issues still need to be addressed. (1) Sophisticated monitoring strategies are needed to capture comprehensive performance data at an acceptable cost. A distributed monitoring framework must be developed on the most up-to-date distributed computing technologies and global cyberinfrastructure. For example, cloud computing and volunteer computing (*i.e.*, a type of distributed computing in which computer owners donate their computing resources to support projects of others temporarily, such as SETI@home [52] and Climate@home [53]) technologies can extend the spatiotemporal coverage of monitoring sites [50]. (2) More metadata, access behaviors and fine-grained monitoring metrics could be recorded or monitored. For example, request error analysis is helpful to locate backend issues. (3) In addition to the basic statistics on performance and accessibility, advanced analysis is needed to reveal the spatiotemporal patterns in service performance (e.g., response time). These features are critical for QoS predication and evaluation [54]. Service selection [49] and server-side optimization [55] could benefit from reliable QoS predication and evaluation results.

These issues motivated us to conduct a thorough resource survey and quality analysis of global WMSs. To capture comprehensive QoS data, a distributed monitoring framework with 27 dispersed monitoring sites was deployed based on public cloud services. The metadata and QoS data of 46,296 WMSs from 72 countries were collected. The imbalance and features were discovered, including the server locations, provider types, supported service versions, popular map subjects, layer spatiotemporal distribution and supported Coordinate Reference System (CRS). We analyzed stability, request error types and potential reasons for error and discovered a power law for the response time distribution. We discuss the spatiotemporal features of response time for individual WMS and show how these discoveries could provide guidelines for service discovery and selection.

## 3. Data Collection and Methodology

The data collection and analysis workflow is illustrated in Figure 1. First, WMSs are discovered using a developed topic crawler. After importing the crawled WMSs into a database, the WMS resource survey and QoS analysis were conducted. The WMS resource survey is based on service metadata (*i.e.*, capabilities documents retrieved from *GetCapabilities* operation). QoS analysis is based on our monitoring result on the two mandatory operations, *GetCapabilities* and *GetMap*. Routine monitoring permits acquisition or updating of long-term QoS behaviors (e.g., stability, performance) and the availability status (whether the URL is not valid any more) of all WMSs, while intensive monitoring captures the spatiotemporal features of QoS for selected services.

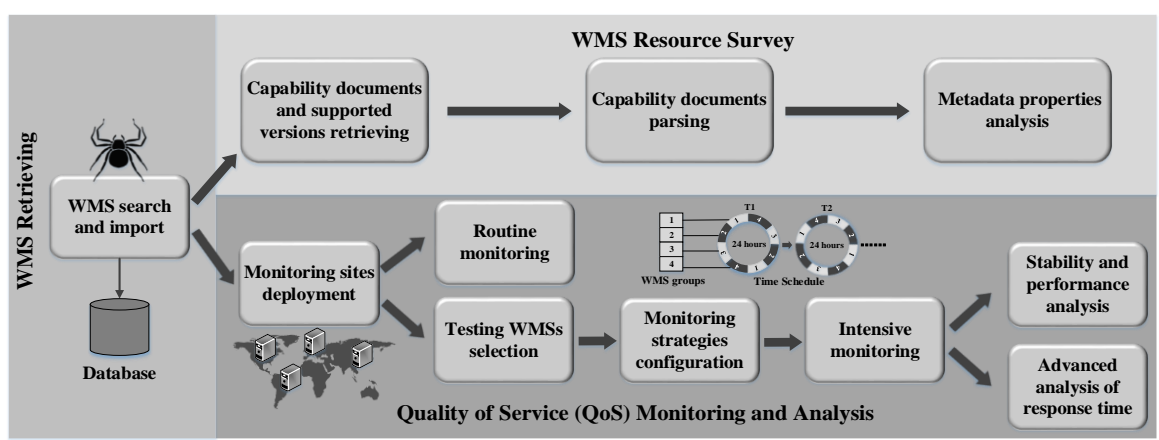

**Figure 1.** Data collection and analysis workflow.

### 3.1. Online Web Map Service Discovery

The WMSs investigated in this research were all collected from the World Wide Web by our active topic crawler [56]. The discovery workflow is illustrated in Figure 2. To collect as many WMSs as possible and to ensure the global coverage of collected WMSs, the crawler adopted a hybrid search strategy that integrates search engine-based search and directed search. Common search engines (e.g., Google and Bing) have powerful crawling and indexing capabilities that capture global web pages. Therefore, a search engine-based search guarantees the breadth of search and, therefore, ensures that our search can reach the regions of the web indexed by common search engines. More to the point, WMS is a type of domain-specific web resource, usually published through geospatial web portals and catalogues, such as the OGC Catalogue Service for the Web (CSW). We used directed search to make a dedicated search of these SDIs using standard APIs and web page crawling. Reputable SDIs (e.g., data.gov [31], GEOSS clearinghouse [30], EuroGEOSS broker [57]) and other geospatial web portals found through search engines were set as seed pages.

In terms of search engine-based search, we developed two methods. The first method searches WMS directly. Keyword-based search (e.g., "WMS", "Web Map Services" or "OGC") or advanced search functions (e.g., in Google Search, we use "service = WMS" as a query constraint for an "*inurl*" statement) provided by the search engine were utilized to retrieve candidate web pages (*i.e.*, seed pages) that may contain WMS URLs. Then, our crawler crawled these web pages recursively to discover WMSs. The second method searches WMS by locating ArcGIS REST Service Directories using keyword-based search (*i.e.*, "ArcGIS REST Service Directory") because many online service directories [58,59] generated by ArcGIS servers [20] provide OGC-compliant geospatial web services. To discover WMSs from such a service directory, the crawler firstly identified real service directory folder pages from retrieved search results by analyzing the HTML structure and content. Then, a dedicated and recursive crawl was conducted on the qualified web pages.

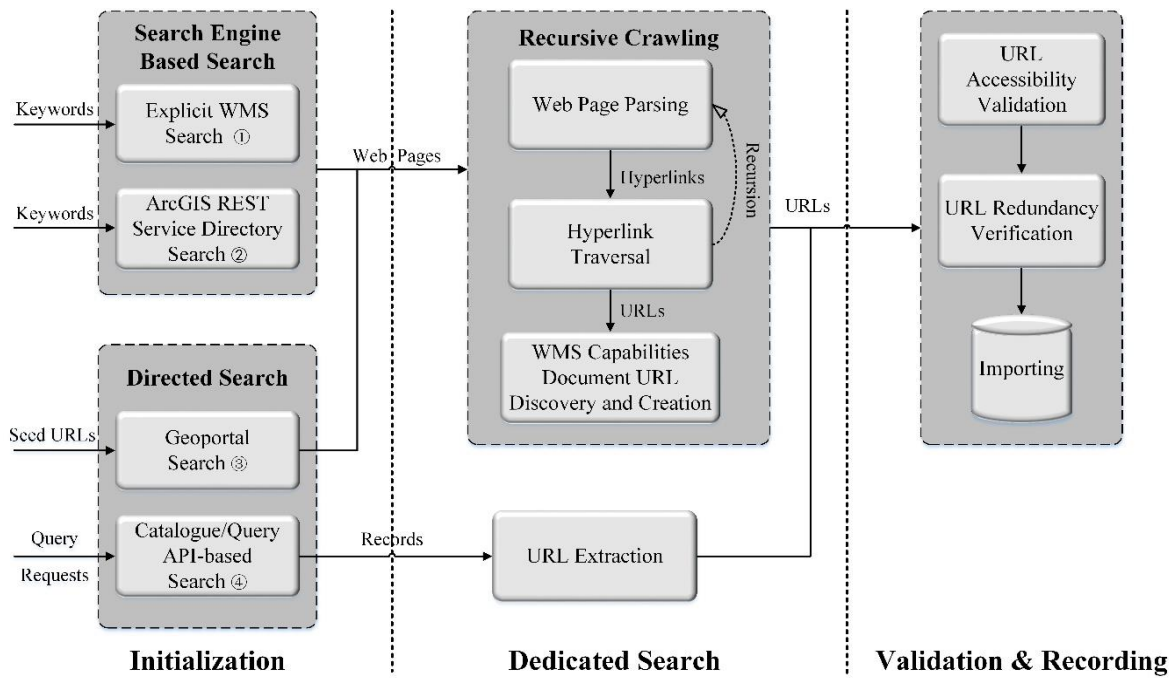

**Figure 2.** Search strategies and discovery workflow of web map services.

The procedure for WMS discovery from a specified web page was treated as a matching and validation process on the WMS capabilities document URLs hidden among all of the HTTP URLs included on the web pages. The HTTP request URL of a WMS capabilities document or its URL prefix is usually posted as a hyperlink or text on web pages explicitly. A canonical HTTP request for the WMS *GetCapabilities* operation contains multiple Key-Value Pairs (KVPs) as essential request parameters, such as "request = GetCapabilities" and "service = WMS". Based on the WMS URL prefix, the capabilities document URL can be easily formed by appending a necessary query parameter. Thus, we used the KVP feature as a criterion to search or form URLs. To avoid unnecessary URL formation processes, only the hyperlinks whose anchor texts and URL syntax that followed specified rules [56] were selected as candidate URL prefixes. Nevertheless, a URL with such a syntax structure cannot guarantee a valid WMS, and a further HTTP request is needed for validation. Our crawler votes for a URL as a valid one only if the payload of the HTTP response is a valid XML document that contains mandatory element tags (e.g., <WMS_Capabilities>). A valid WMS URL was added into the database if it did not replicate a URL already present in the database.

*3.2. Distributed Monitoring Framework and Strategy*

To collect QoS monitoring data from geographically-dispersed monitoring sites, we developed a distributed monitoring framework that consists of three components (Figure 3). (1) The data collector collects service metadata and real-time performance data of WMSs using a group of monitoring sites. Each monitoring site monitors the assigned WMSs in parallel using multithreading technologies. A monitoring manager configures and coordinates the monitoring tasks of all monitoring sites and is in charge of data management. (2) The data access service provides both the RESTful and Simple Object Access Protocol (SOAP)-based web service APIs to retrieve monitoring data (e.g., monitoring sites, service metadata and historical performance in a given time period). (3) A web portal, developed based on our previous research [60], visualizes the layers, service performance, as well as the spatial distribution of the monitored WMSs and monitoring sites using maps and charts.

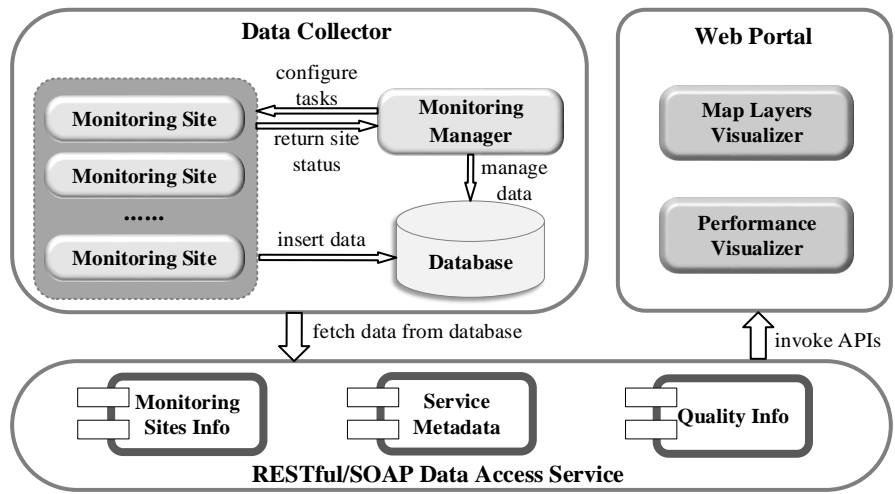

**Figure 3.** The architecture of the monitoring framework.

Both routine monitoring and intensive monitoring were based on our distributed monitoring framework, but the monitoring strategies were different. Routine monitoring utilized several fixed monitoring sites to guarantee that the available status of the WMSs was not impacted by single monitoring site connectivity. The two mandatory operations of each WMS were both tested on a weekly basis, one request for each operation from each routine monitoring site per week. In contrast, to capture performance differences and investigate spatiotemporal patterns, intensive monitoring used more monitoring sites from different locations, and the monitoring time intervals are more intensive and adjustable. In practice, to provide stable accessibility, some WMS providers may setup license policies [61] to prohibit excessively frequent accesses from a single IP address. We respected these policies and conducted distributed round-the-clock monitoring. Multiple monitoring sites were deployed on dispersed locations around the world. They worked collaboratively to capture the performance difference caused by the diversity of spatial positions where users are accessing the services. In the temporal dimension, WMSs were divided in groups and monitored periodically, according to a configurable schedule. The monitoring schedule helped us to investigate the daily pattern in performance and to avoid being blocked by the WMS servers due to excessively frequent requests. Meanwhile, by dividing WMSs into groups, the accumulated response delay within a group could be controlled and the monitoring time interval for a single WMS could be guaranteed. After iterative monitoring and merging multiple day monitoring data into one day, we obtained intensive daily monitoring records with almost equal time intervals for each mandatory operation of a selected WMS, *i.e.*, a record around every five minutes.

For a consistent and comparable test, we created testing rules for the two operations. During *GetMap* operation monitoring, the first named layer [2] of the WMS was requested for testing. We restricted the output image to be a 200-pixel height, 400-pixel width and in PNG format. If this format was not supported, JPEG or other formats were specified accordingly. Although output maps may be distorted as the geographic extent (*i.e.*, bounding box) varies, the data volume can be roughly controlled, as well as the data transfer cost. Meanwhile, to eliminate the impact of differences in server-side processing procedures, we limited each request to a single layer, and requests that combined multiple layers simultaneously were excluded. However, a request upon cascading parent layers that contain child layers was allowed, since a WMS treats a parent layer as a single layer in terms of *GetMap* requests. For *GetCapabilities* operation monitoring, as WMS may support multiple service versions, testing requests that do not specify the optional parameter "version" were used as the measurement to obtain QoS for the default version.

*3.3. Survey Data Collection*

From September 2014 to November 2015, 27 public cloud-based (Windows Azure) monitoring sites dispersed at 13 locations were deployed on four continents and in seven countries (Figure 4) in total. All monitoring sites were employed with the same virtual machine configuration, including the network, computing resources and operating system to avoid any impact on monitoring results as much as possible. Four monitoring sites located in U.S. (Bristow, VA; Redmond, WA), Ireland (Dublin) and China (Hong Kong) were constantly employed for routine monitoring from September 2014 to November 2015. The remaining 23 monitoring sites from 12 locations were set up for intensive monitoring from 23 August 2015 to 3 October 2015. At any time point during intensive monitoring, 12 sites from 12 different locations were selected to conduct monitoring simultaneously; other sites from replicated locations were used as alternatives.

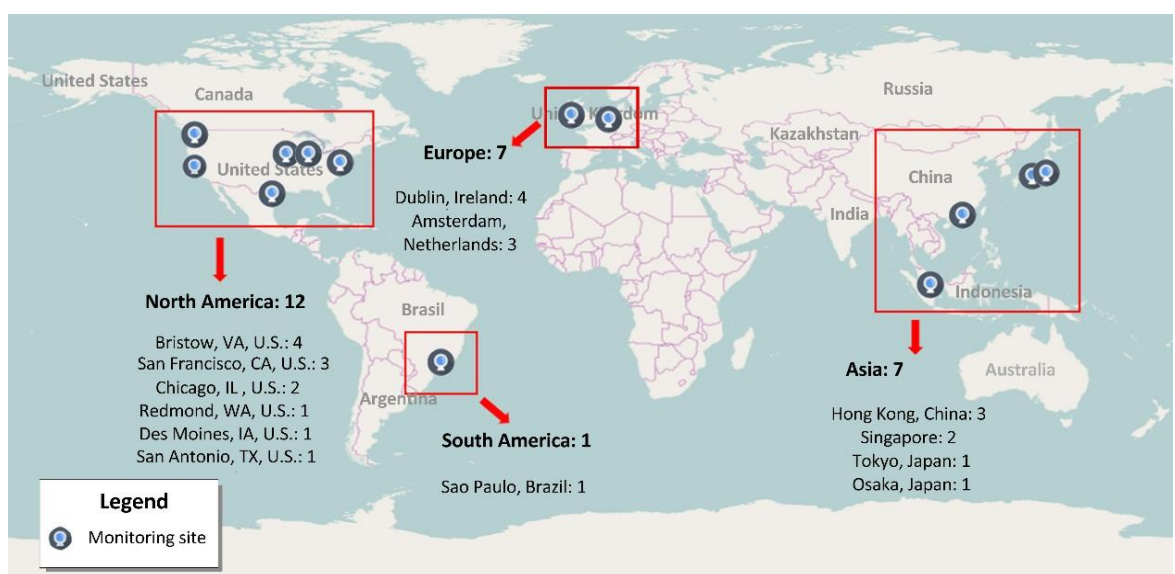

**Figure 4.** The geo-locations of monitoring sites.

Over more than one year of routine monitoring, 46,296 WMSs (from 72 countries and six continents) collected by our web crawler were constantly monitored. Among these services, 41,703 WMSs in total were valid and contained 318,102 layers. Since the total number of WMSs was too large to conduct a comprehensive test, we selected 1210 WMSs to be intensively monitored for 42 days. To avoid the impact of access overload, the number of WMSs from a single provider, the same domain name or IP address was strictly limited to at most five during WMS selection. The global distribution of the selected WMSs was considered, as well. The *GetCapabilities* test was based on these 1210 WMSs. Among these WMSs, 876 WMSs were actually valid for access. We selected them to conduct further *GetMap* tests. Table 1 lists the amount of WMSs from each continent in testing. In keeping with the round-the-clock monitoring strategy as described in Section 3.2, we finished a monitoring cycle every six days. From each monitoring site location, we collected 2016 *GetCapabilities* QoS records (*i.e.*, 48 records per day) for each WMS selected for *GetCapabilities* test and 2016 *GetMap* QoS records for each WMSs selected for *GetMap* test.

**Table 1.** The WMSs selected in intensive monitoring.

| WMSs | North America | Europe | Asia | South America | Africa | Oceania | Total |
|---|---|---|---|---|---|---|---|
| *GetCapabilities* test | 718 | 428 | 11 | 16 | 3 | 34 | 1210 |
| *GetMap* test | 526 | 306 | 4 | 15 | 0 | 25 | 876 |

To conduct a comprehensive investigation, we collected more metadata fields (e.g., contact information, CRS, spatial coverages of layers) and fine-grained quality information than the research reported in the literature [42–45,49,50]. Response time, error type, size of response message, download speed and relevant monitoring site were recorded from each monitoring request. Response time is affected by the latency on the client, network and server side. To analyze the key impact factors of WMS response times in the future, we obtained fine-grained response time costs for different phases in the interaction. The total response time in a complete HTTP request-and-response operation was decomposed into different time spans using the cURL command line tool [62]. The attributes of these time spans include Domain Name System (DNS) parsing time; connecting time; total time for sending a request and server processing; and the data transfer time. In addition, the average response time and the successability for each WMS were updated periodically as quality metrics.

## 4. Data Analysis

### *4.1. Global WMS Resource Survey*

In the resources survey, primary metadata properties, including the title, abstract, keywords, coordinate reference system, version and provider, were extracted from the WMS capability documents to investigate service usage and resource distribution.

4.1.1. Server Location and Provider Type

The geo-location of monitored public WMSs suggests that North America (especially the U.S.) and Europe have most abundant WMS resources (Figures 5 and 6a). In the U.S. and Europe, Open Government Data initiatives promote the accessibility and re-use of official data and information by citizens, communities and developers via open development repositories. Spatial data infrastructures (SDIs) developed in these regions are widely used by cross-domain users all over the world for geospatial resources discovery and sharing. For examples, data.gov [31] integrates the U.S. government's open data to build a one-stop data catalog, and the INSPIRE geoportal [63,64] is a European Union (EU) SDI to facilitate the sharing of environmental spatial information for public use. As a result, the global accessibility of geospatial resources published in the U.S. and Europe is significantly enhanced, and online service exploration becomes much easier. The geo-location distribution of WMSs does not necessarily mean there are more map resources in the U.S. and Europe. This distribution indicates that OGC service standards are more widely adopted for publishing open data in those places. Consequently, these resources are more easily reached by global public users through geoportals and search engines.

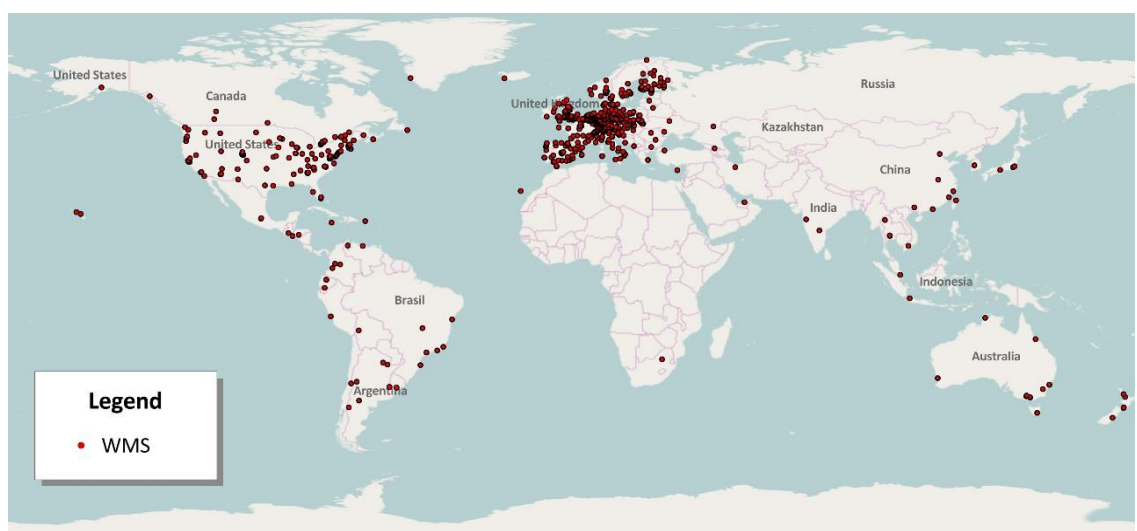

**Figure 5.** The geo-location of monitored public WMSs.

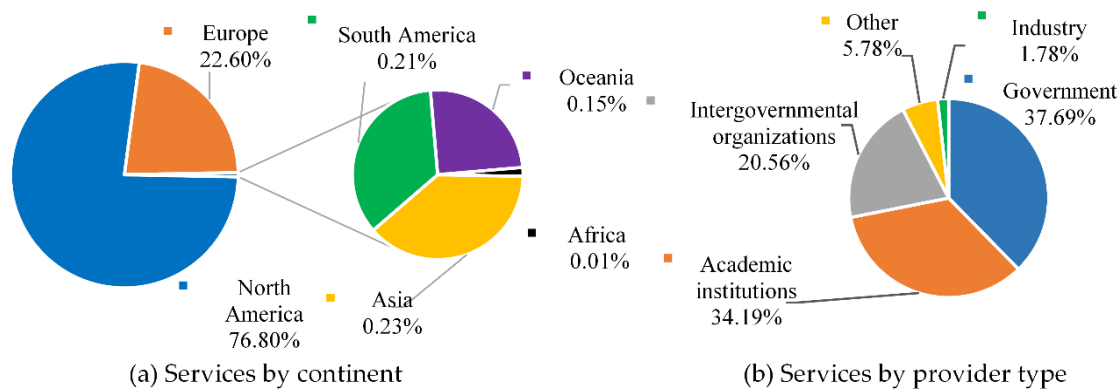

(a) Services by continent　　　　　　　　　　(b) Services by provider type

**Figure 6.** The regional distribution and types of monitored public WMS providers; (**a**) while most services are in North America and Europe, among the remaining 0.6%, Asia contains 0.23%; (**b**) providers by sector; government has the largest proportion (37.69%), followed by academic institutions (34.19%), while industry has the smallest proportion (1.78%).

From an analysis of the 13,352 WMSs from 989 providers that include provider information, such as organization tags in their capability documents, we found that the server locations and provider types of WMSs are imbalanced. The WMS standard draws wider support from governments, academic institutions and intergovernmental organizations to share non-profit (e.g., public welfare and resources) geospatial data (Figure 6b) than industries. Providing public data services is one of the duties of governments and intergovernmental organizations in order to benefit society. Therefore, they actively publish data related to Societal Benefit Areas (SBAs) [65] and other public interest areas using open standards. Academic institutions also play a major role in proposing and using open standards to facilitate scientific data sharing. In contrast, services provided by industry only represent a small proportion (1.78%), because WMS is an open standard for geospatial data access rather than a commercial-level data exchange protocol.

By counting the number of WMSs published by each of the identified 989 providers, we found that the top ten service providers (1%) of analyzed WMSs published 7367 WMSs (55.18%, 13,352 in total), but the number of WMSs they offered varies significantly (Table 2). The Earth Data Analysis Center (EDAC) of the University of New Mexico and National Oceanic and Atmospheric Administration (NOAA) make the largest contribution. EDAC hosted and managed the New Mexico Resource Geographic information System (RGIS) [66] for over 23 years specifically to share statewide public geospatial data. The data are managed by the RGIS data repository and published through OGC-compliant web services. NOAA, as well, provides a huge amount of oceanic- and atmospheric-related data through WMS. The imbalanced service contribution among providers follows a power law [67], as shown in Figure 7, and reflects the differences in geospatial resource possession, data sharing policies and effectiveness.

**Table 2.** Top ten service providers in the monitored WMSs.

| Provider Name | Service Amount | Provider Type |
|---|---|---|
| Earth Data Analysis Center (University of New Mexico) | 3202 | Academic institution |
| National Oceanic and Atmospheric Administration (NOAA) | 2958 | Government |
| Food and agriculture organization of the United Nations (FAO) | 295 | Intergovernmental organization |
| Vlaamse Overheid (Flemish Government) | 262 | Government |
| Senatsverwaltung für Stadtentwicklung und Umwelt (Senate Department for Urban Development and the Environment of Berlin) | 169 | Government |
| United States Geological Survey (USGS) | 133 | Government |
| Landesamt fuer Geologie und Bergbau, Rheinland-Pfalz (Department for Geology and Mining of Rheinland-Pfalz) | 101 | Government |
| Arizona Geological Survey | 98 | Government |
| Illinois State Geological Survey | 91 | Government |
| Kansas Biological Survey | 58 | Government |

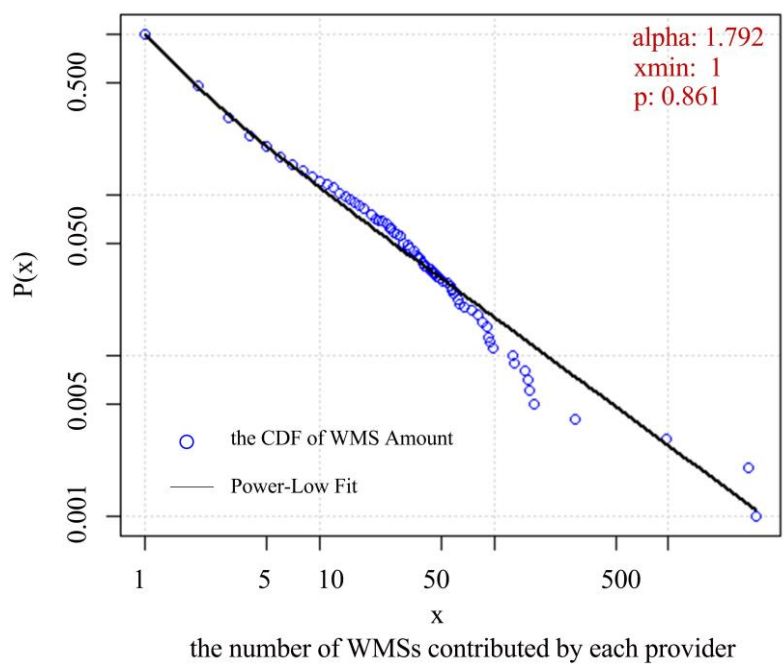

**Figure 7.** Log-log plot of the Cumulative Density Functions (CDF) p(x) for the number of WMSs contributed by each provider according to a discrete power law with α = 1.792 and $x_{min}$ = 1.

We further summarized the prevalent software used to publish WMSs by analyzing WMS Uniform Resource Locators (URL) and capabilities documents. Among a total of 46,296 WMSs, there were 3484, 1987 and 515 WMSs published by ArcGIS Server, GeoServer and MapServer, respectively; the publishing software for the remainder is unknown. Government agencies published 1721 of the ArcGIS-based WMSs. In contrast, only 229 WMSs from government agencies were published using GeoServer and MapServer. The two open-source servers have academic origins, hence more widely adopted in the academic sector than the public sector. Both commercial and open-source geospatial software contribute to WMS publishing, but open-source tools are more popular in academia. Commercial software, such as ArcGIS, is more likely to be used by government agencies given the governmental contracts with software companies.

4.1.2. Popular Map Subjects

Investigating the map subject matter of global WMSs can benefit global cross-disciplinary users in discovering and selecting map resources, as well as increasing our understanding of open data sharing policies, public issues, and trends in the Earth sciences and other disciplines. Although the languages used to describe WMS metadata vary (e.g., English, German, French, Dutch, Italian, Norwegian, Swedish and Chinese), according to our investigation, around 87.85% of the valid WMSs (36,638) contained English words in their service descriptions, and 82.57% of the layers contained keyword fields that were described in English. Subsequently, we studied the top map layer subjects based purely on English keyword to reduce the complexity of data analysis. We extracted a group of English keywords for each map layer from layer description fields (e.g., title, abstract, keywords) of capability documents. The stop words, replicated words and words that were not nouns were eliminated during extraction. Then, we calculated and sorted the occurrence frequency of keywords among all map layers. Figure 8 shows the top English language keywords that had the highest word frequencies. According to the GEOSS Societal Benefit Areas (SBAs) [65] and INSPIRE directive [63], we analyzed obtained keywords and found that the following subjects related to the natural environment and resources appear most frequently: geology, climate, energy, land cover, water, biodiversity, agriculture and ecosystem. Given the explosive growth in Earth observation technologies over past several decades, governments, academic institutions and non-profit organizations have collected and processed a large amount of geographic data about natural phenomena. Many of the data were published using standard OGC services and form the majority of open map resources. For example, the INSPIRE directive is primarily oriented to share spatial environmental information among public sector providers [63].

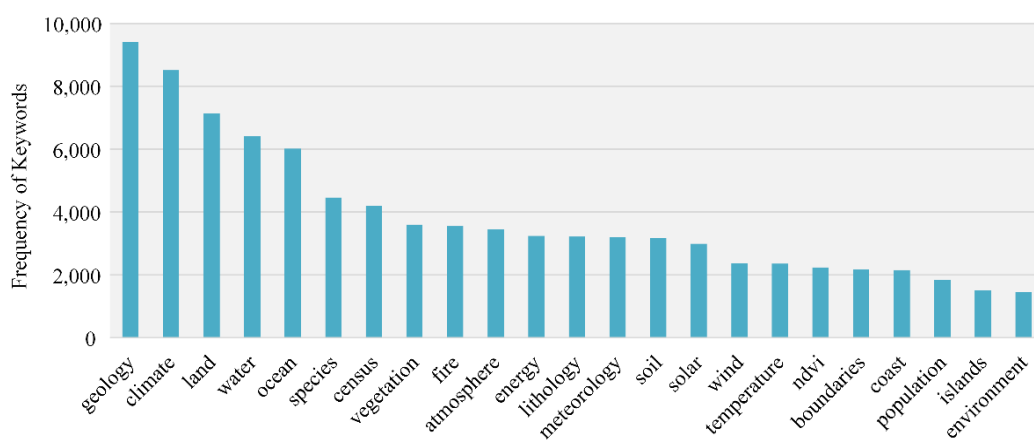

**Figure 8.** Top English language keywords found in service descriptions for map layers.

The big data era has arrived and given the advancement of location-aware technologies; social sensing is becoming easier and less expensive. Accordingly, social and economic activity-related data are dramatically increasing. The Socioeconomic Data and Applications Center of Columbia University (SEDAC) published a WMS [68] to represent the spatiotemporal distribution of population size, population density and education degree in America. The National Geomatics Center of China (NGCC) published thematic maps of Chinese employment and wages in 2013 as a WMS [69]. It expected that online maps with social behaviors and economic activities as primary subjects will be exploding exponentially in the near feature.

4.1.3. Spatial Coverages of Map Layers

By analyzing the geographic extent (*i.e.*, bounding box) of 318,102 map layers (Figure 9), we found that continents are covered more intensively than the oceans, except for Antarctica, and the Northern Hemisphere has more coverage than the Southern Hemisphere. Many of the layers have a global extent (more than 25,000 layers). This phenomenon reveals our global Earth observation behaviors and also reflects human activities in space, to some degree. The differences in the richness in open geospatial data also reflect the differences in data sharing policies at the country and regional levels. More specifically, spatial coverages of map layers are concentrated in North America and Europe, especially over the mainland of U.S. and Europe, and Northern Africa has the second largest number of coverages, after North America and Europe. However, due to information security and data policy issues, geospatial resources are strictly controlled in some countries. Geospatial data are often shared internally and between governmental agencies through hardcopy or secured private networks rather than by publishing them through standardized web services for public use.

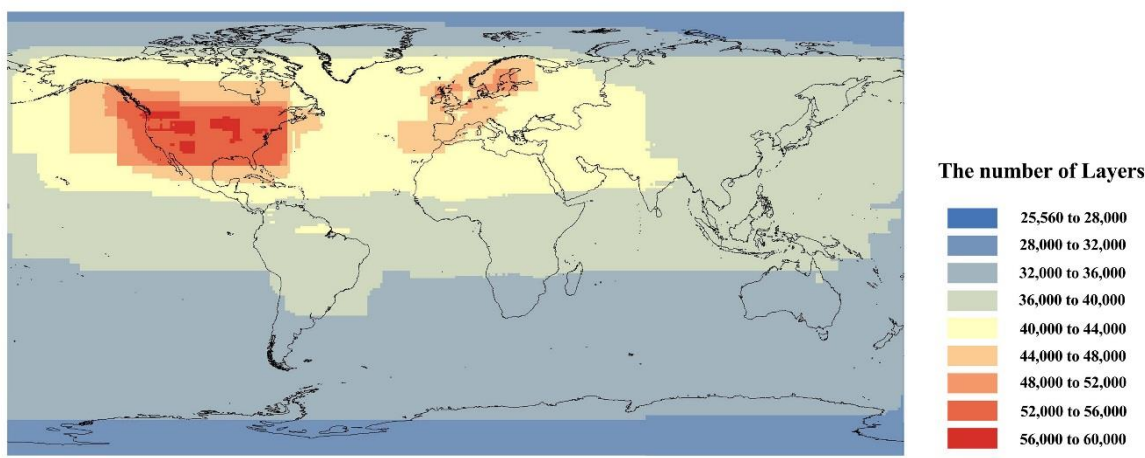

**Figure 9.** The spatial coverages of Map Layers.

4.1.4. Yearly Distribution of Map Layers and WMSs with Current Map Layers

The data collection time is a measure of the timeliness of geospatial data (*i.e.*, data currency). Among all 318,102 layers, 62,925 layers from 4587 WMSs contain data collection times in their capability documents. For example, the SEDAC GeoServer WMS [68] embeds the data collection time directly in the layer name field (e.g., "Population Density 2000"), and the NASA Earth observation WMS [70] describes an extended time dimension in each layer (e.g., "2015-01-01/2015-08-29/P8D"). Using such information, we can estimate approximately the update status of map layers or service metadata for a WMS. As data become more open, the map resources collected each year are increasing in number. At the same time, most of these data published using WMSs are up-to-date data in general, not archived historical data (as shown in Figure 10).

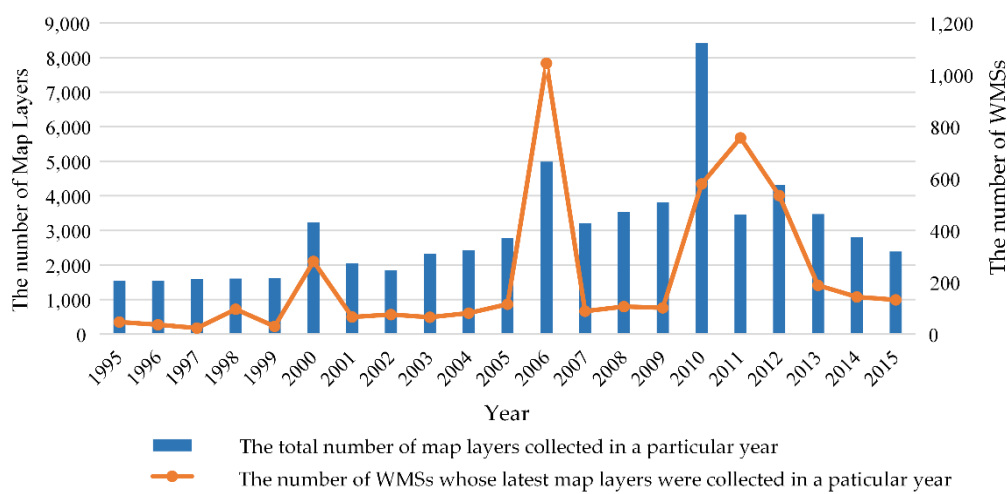

**Figure 10.** Yearly distribution of map layers and WMSs with current map layers.

Figure 10 illustrates that the significant increase in WMSs and map layers in 2000 and 2006 was strongly associated with the release of the WMS standards Version 1.0.0 and Version 1.3.0, respectively. The release of the ESRI RESTful Service APIs and successive updates, which are compliant with the WMS standard, caused another significant increase, starting in 2010. Moreover, INSPIRE required member states to provide discovery and view services (*i.e.*, WMS and WMTS) in 2011 at the latest, which may also have promoted the deployment and updating of WMSs. These trends reveal that new technologies, standards and software products are adopted relatively rapidly in the geoinformation domain. However, the enthusiasm for exploring new technologies is hard to maintain when it comes to routine service maintenance. Many of the existing WMSs seldom add new layers or update layer metadata after deployment. Thus, the data collection time for the latest layers in 4124 WMSs among the 4587 WMSs studied (89.91%) is earlier than the year 2013. This may affect the quality of map resources found in online WMSs.

4.1.5. Supported Coordinate Reference Systems and Service Versions

Among the 318,102 layers examined, various ellipsoidal CRSs and projected CRSs were supported. Ellipsoidal CRSs obtained 97.57% of the support. Web Mercator, Universal Transverse Mercator, Antarctic stereographic and Albers are the most used projections (Table 3). Web Mercator obtained 76.39% of the support. Maps in this projection can be conveniently visualized on the web, since it is easy to conduct map splitting and seamless splicing. Meanwhile, Web Mercator guarantees the correctness of direction and relative position on maps. Therefore, it is widely adopted, and many online public map services use this projection. Some layers support the Antarctic stereographic projection or Albers projection due to the geographic extent of the maps or because of their specific application needs. For example, administrative region maps may need equal-area projection.

**Table 3.** Common map projections of widely-supported projected Coordinate Reference Systems (CRSs).

| Map Projection | Layers Amount and Percentage | CRS Sample |
|:---:|:---:|:---:|
| Web Mercator | 241,779 (76.39%) | EPSG:3857, EPSG:102100 |
| Universal Transverse Mercator | 163,092 (51.52%) | EPSG:26919, EPSG:32633 |
| Antarctic stereographic projection | 122,399 (38.67%) | EPSG:3031 |
| Albers projection | 119,024 (37.61%) | EPSG:3005 |

Among 41,703 valid WMSs, there were 9920, 10,122, 41,325 and 40,861 WMSs supporting Versions 1.0.0, 1.1.0, 1.1.1 and 1.3.0, respectively. Therefore, most of the WMSs were implemented with geospatial software instances compliant with the latest WMS standard Version 1.3.0, released in 2006 and compatible with Version 1.1.0. In Version 1.3.0, the access interfaces were unified, and the data structure of response messages was refined. Therefore, the version is much more mature than previous versions, contributing to the high adoption rate. Meanwhile, after several years of popularization, the recognition of OGC standards was significantly promoted among users. More open-source and commercial GIS software developers began to support the latest OGC standards. It was also noted that most of the WMSs with Version 1.3.0 are downwardly compatible with Version 1.1.1, but only around 1/4 of WMSs are downwardly compatible with the earlier versions (1.0.0 and 1.1.0).

*4.2. Stability and Performance Analysis*

Stability and performance are two essential service-level measurement quality factors for web services [71] and software entities [72]. Stability measures the reliability and the maintainability of a software entity by investigating the runtime robustness, while performance measures runtime efficiency. Evaluating these two factors is significant in that it may provide guidelines for WMS selection and server-side improvements for service consumers and providers, respectively. In this section, based on the intensive monitoring result of 1210 WMSs (listed in Table 1), we analyze the overall status of the two factors using selected metrics.

4.2.1. Stability Analysis

To analyze the stability, we investigated accessibility, successability and error types based on two mandatory operations. Accessibility represents the probability that a web service operation is accessible (receiving acknowledgement or response message) while the service is available. Successability is the ratio of successful responses to all requests in a time period and measures the ability to correctly respond to user requests. Error type describes briefly the reason why a request failed. We categorized accessibility into three types in this research: always accessible, temporally inaccessible and constantly inaccessible. Table 4 shows that many WMSs are constantly inaccessible (*i.e.*, invalid WMSs), since these are unavailable or they may have changed URLs. Meanwhile, WMS operations identified as temporally inaccessible were a nontrivial problem and due to network and service maintenance issues. Since we cannot detect or be informed of the maintenance times of all WMSs, maintenance time was not excluded from accessibilities analysis. The accessibility of *GetMap* is worse than *GetCapabilities* even for the valid WMSs. Only around 1/5 of layer *GetMap* operations are constantly accessible. The successability histograms for *GetCapabilities* and *GetMap* seen in Figure 11 also indicate that the successability of *GetCapabilities* is higher than *GetMap* for the valid WMSs.

**Table 4.** Accessibility of *GetCapabilities* and *GetMap* operations.

| Accessibility | *GetCapabilities* for All Selected WMSs | *GetMap* for Valid WMSs |
| --- | --- | --- |
| Constantly inaccessible | 27.60% | 17.21% |
| Temporally inaccessible | 13.64% | 61.43% |
| Always accessible | 58.76% | 21.36% |

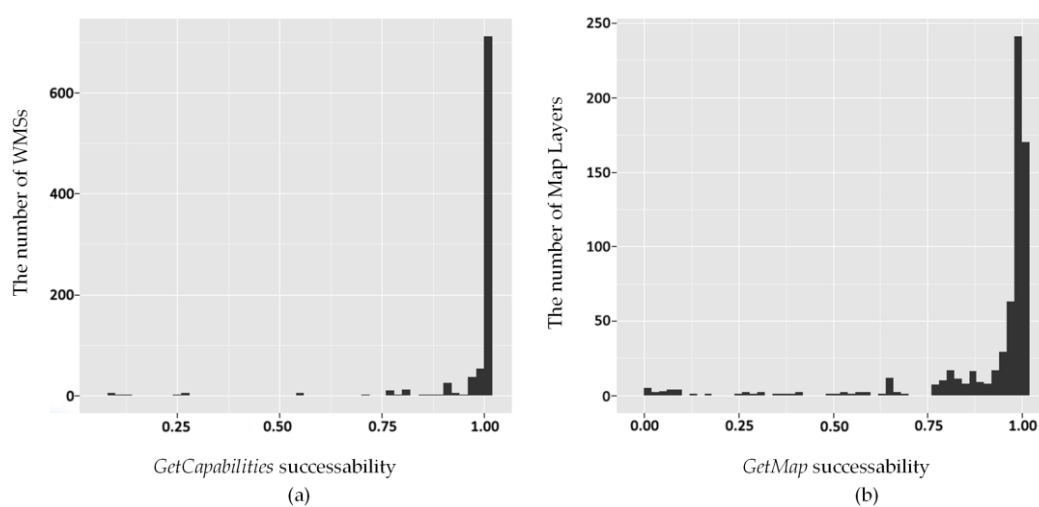

**Figure 11.** Successability histograms of the two mandatory operations for valid WMSs: (**a**) *GetCapabilities*; (**b**) *GetMap*.

To analyze operation errors, we classified various detailed error types into two categories. The server access error represents the errors that occurred when connecting to the servers, e.g., being unable to connect the host, a time-out or no response from the server. Request processing errors happen during server-side processing after successfully connecting to the server, e.g., semantic error of request, server refusing to execute the request or server overload. From Table 5, we can see that more errors were caused by request processing errors for *GetMap*. On the one hand, the processing of *GetMap* operations is relatively more complex than *GetCapabilities*. *GetMap* needs to load geospatial data and conducts requisite geoprocessing (e.g., subsetting, transformation and rendering) to generate maps, while *GetCapabilities* only responds to a metadata document that can be generated in advance. On the other hand, incorrect or outdated layer descriptions in the capabilities document are another reason for request processing errors. We can conclude that stability varies significantly for services and operations. Accessibility to the metadata cannot grantee the accessibility to the map layers. Therefore, the stability of WMS servers and metadata timeliness maintenance are big issues and need to be further addressed by service providers.

Table 5. Error types in *GetCapabilities* and *GetMap* operations.

| Error Type | *GetCapabilities* | *GetMap* for Valid WMSs |
|---|---|---|
| Request processing error | 61.64% | 76.42% |
| Server access error | 38.36% | 23.58% |

4.2.2. The Power Laws in Response Times

Response time, maximum throughput and computing resource occupation are three common measurement of performance. In our research, we selected response time as the performance measurement because testing maximum throughput requires intensive concurrent requests and may result in request rejection by service providers, while computing resource occupation measurements are hard to obtain. We investigated the overall trends in the response times of all tested WMSs by analyzing the minimum, average and maximum response times of the two mandatory operations recorded for each valid WMS among all successful responses from all monitoring sites. Figure 12 indicates that the response time of the two mandatory operations for valid WMSs, in general, must obey power laws. Most WMSs respond rapidly, but a few have very long response times. The numerical differences between the minimum and the maximum response times are illustrated in Figure 13 reflecting instability in WMS performance. The vertical red lines in the bar charts for average response times indicate that a majority of valid WMSs (more than 80%) can respond to user requests within three seconds in most cases. In contrast, our previous investigations show that less than 40% of WMS *GetCapabilities* and *GetMap* operations responded within eight seconds [48,49]. Thus, the overall response time of WMSs was significantly reduced, reflecting improvement in WMSs software and hardware environments, as well as upgrades in the global network, to some extent.

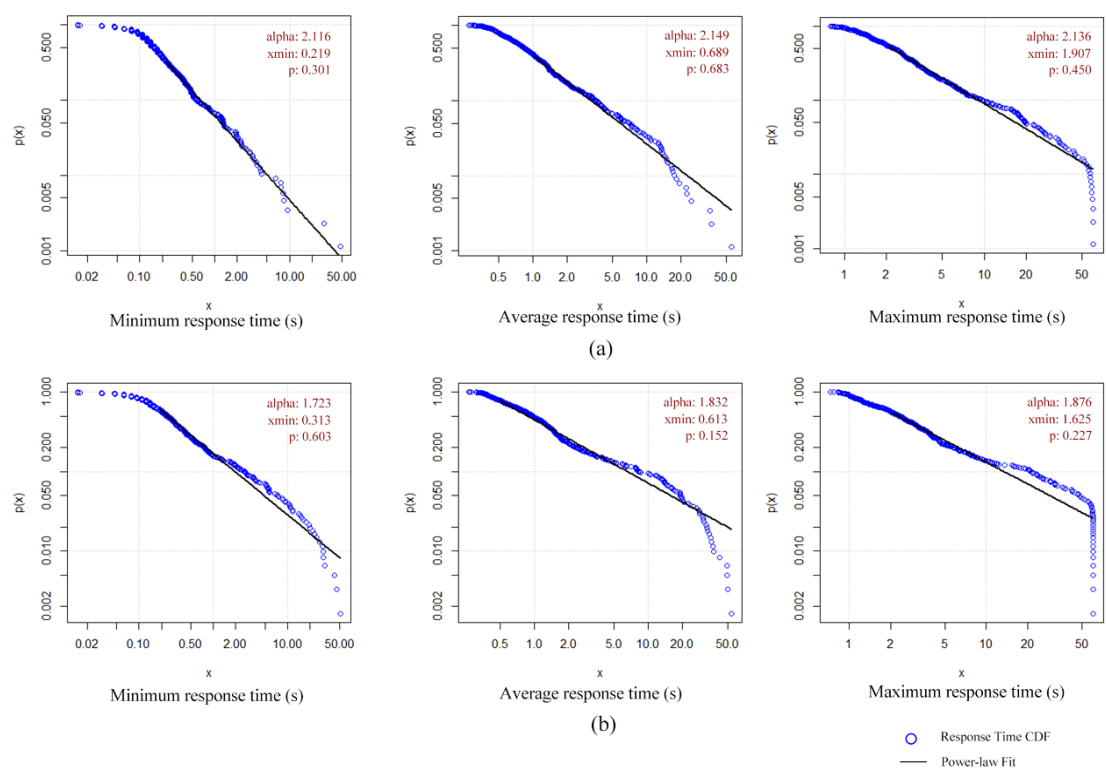

**Figure 12.** Log-log plot of the CDFs P(x) for minimum, average and maximum response times of (**a**) *GetCapabilities* operations and (**b**) *GetMap* operations for selected WMSs reflecting continuous power laws; a Kolmogorov–Smirnov test ($p > 0.05$) indicated that the original data were likely to be drawn from the fitted power-law distribution. The plots show a sharp change in the upper boundaries of the response time dropping significantly at about 60 s, especially in the maximum response time, because the maximum timeout was set to 60 s during monitoring.

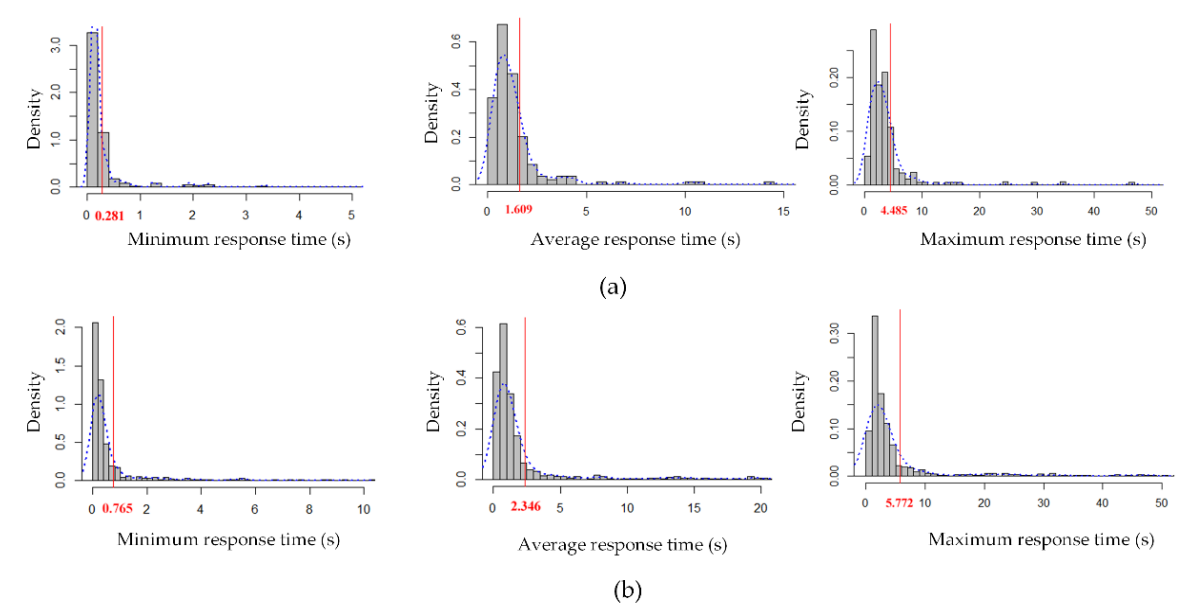

**Figure 13.** Unit area histograms for the minimum, average and maximum response times of (**a**) *GetCapabilities* operations and (**b**) *GetMap* operations for selected WMSs; the vertical red line in each chart divides the WMSs into 80% and 20% proportions, respectively. The density, calculated as *frequency*/(*total_frequency*bin_width*), shows the proportions of WMSs for per unit of response time.

4.2.3. Spatiotemporal Characteristics of Response Times

The response time of a web service is affected by various factors. Among these factors, the network connectivity between service providers and users, instantaneous network condition, as well as the concurrency pattern of global users can be generalized as spatiotemporal access factors. In this section, we explore the relation between spatiotemporal access factors and the response time.

(1) Spatial characteristics

Response time is significantly impacted by network connectivity, and the connectivity of a network in cyber space relies on the establishment of network equipment and their linkages in geographic space. Therefore, the association between spatial distance and response time is inherent. We found that 60.27% of valid WMSs (528 of 876 in total) obtained the shortest average response times from their closest monitoring site locations globally. At the continental level, this trend was more apparent (Table 6). Most of the WMSs tend to get the shortest response time from the monitoring sites in the same regions as the server locations, except in the case of South America. We also calculated the linear regression coefficient of determination ($R^2$) of the response time for individual WMSs using the average response times at each monitoring location for multiple locations including both the public cloud-based sites and local sites. The average $R^2$ was 74.08% for all 876 WMSs. Figure 14 reveals a positive correlation between average response time and the spatial distance from monitoring site to server. The scatter plot in Figure 14a is for all 876 valid WMSs. We can see a positive correlation, but the graph also shows data heterogeneity, since response time is impacted by many other factors, such as the response data volume, server performance, network bandwidth and the distribution of global optical cables. The impact from these factors was partially reduced by selecting 393 WMSs located in the U.S. whose capabilities documents were less than 1 MB and had an average response time of less than 2 s. The positive correlation becomes more visible in Figure 14b. Furthermore, the shorter the distance, the smaller the variance in response times, as response time becomes more stable when the uncertainty of a network is reduced. Although the response time is impacted by many factors and hard to predict precisely, we make the following suggestions. From the perspective of service selection, a WMS that has a closer geographic distance to users may have a higher priority among services with comparable functionalities and map resources. From the perspective of performance optimization, a map server should be deployed as close as possible to potential users. Cloud computing can be utilized to achieve dynamic spatiotemporal deployment of servers, and the state-of-the-art site selection algorithms could be developed to improve performance.

**Table 6.** The percentage of WMSs from each continent that yielded the shortest average response times from the monitoring sites on each continent.

| Monitor Location | WMSs by Continents (Service Amount) | | | | |
|---|---|---|---|---|---|
| | North America (526) | Europe (306) | Asia (4) | South America (15) | Oceania (25) |
| North America | 91.25% | 3.59% | 0 | 60% | 4% |
| Europe | 3.04% | 95.10% | 0 | 0 | 0 |
| Asia | 3.80% | 0.65% | 100% | 6.67% | 96% |
| South America | 1.90% | 0.65% | 0 | 33.33% | 0 |

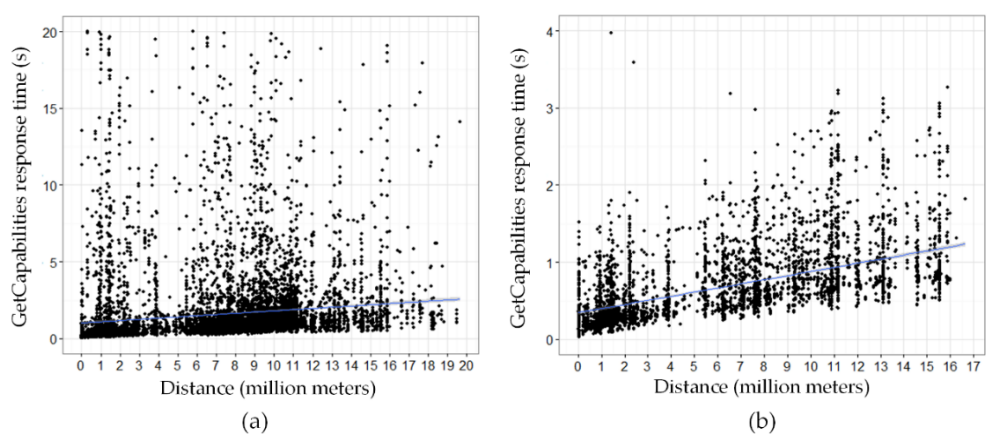

**Figure 14.** Correlation between response times and spatial distance from monitoring sites to WMS servers. (**a**) The 876 valid WMSs and (**b**) the selected 393 WMSs located in the U.S. whose capabilities documents were less than 1 MB with average response times less than two seconds.

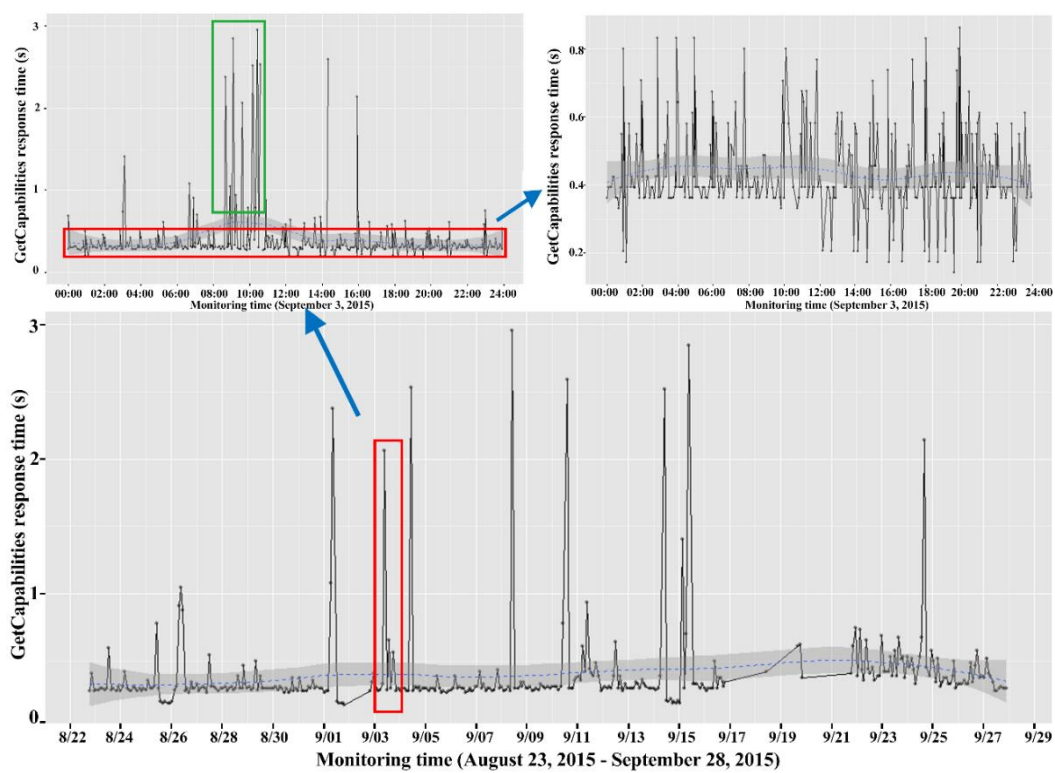

**Figure 15.** Time series characteristics of WMS response times.

(2) Time series characteristics

The monthly response time series for a WMS from a single monitoring site is steady in general but synthesizes trends with few prominent random fluctuations and many small local variations. Within the 24 h of a day, a time series shows local fluctuations. As shown in Figure 15, for a WMS [73] provided by Arizona Geological Survey, there is a set of intensive peaks between 8:00 a.m. to 11:00 a.m. in the local time of the service, while the response time fluctuates slightly during the night. This phenomenon reveals the local network status, as well as concurrent accesses to a WMS during a certain time period, to some extent. The local network traffic and user concurrency from the same or neighboring time zones to the WMS server are relatively small in the nighttime, but increase sharply during the daytime. Concurrent access to the WMS generates server-side load pressure. The fluctuation is more violent for the WMSs with a longer average response time, since the network condition has a huge impact on the stability of the response time [50].

## 5. Conclusions and Future Work

### 5.1. Conclusions

We conducted a comprehensive WMS resource survey and quality analysis for global WMSs based on a proposed distributed monitoring framework. Based on a WMS resource survey of 41,703 valid WMSs, we found that the WMS standard is widely adopted with an imbalanced distribution. More specifically, (1) the providers and server locations are extremely imbalanced. A few providers provided a large amount of public WMSs. WMSs are readily adopted by governments, academic institutions and intergovernmental organizations. In contrast, the contribution from companies is relatively small, since WMS is an open standard for geospatial data access rather than a commercial-level data exchange protocol. Public WMSs also have a skewed spatial distribution due to data policy issues and the imbalanced development of SDIs. Specifically, North America (especially the U.S.) and Europe contributed most of the public WMSs (around 99%). (2) Map resources are abundant, but also disproportionate. The natural environment and resources are the dominant map subjects. North America and Europe have the most concentrated layer coverages. Most of the map data were collected since 2000 when the first version of WMS 1.0.0 was released. (3) The ellipsoidal coordinate system is supported by most of the WMSs, and the Web Mercator projection is widely supported. Most WMSs are published based on the latest Version 1.3.0 and compatible with Version 1.1.1, but the downward compatibility with the old versions (*i.e.*, 1.0.0 and 1.1.0) is deficient.

From the quality analysis, we found that the quality monitoring, evaluation and optimization are imperative and of critical importance for WMS. (1) The quality of the WMSs varies on services, operations and request parameters. Plenty of WMSs are inaccessible due to invalid URLs. *GetMap* has weak stability and accessibility when compared to *GetCapabilities* due to the relatively complicated processing in the *GetMap* operation and incorrect layer descriptions. Request processing errors are the major factor that causes request failures. (2) The response times of all valid WMSs obey power laws. The majority of WMSs can respond rapidly (within three seconds) generally, while a small number of them have long response times. However, when compared to contemporary commercial online map services, the large interval between the average and the maximum response times reveals ubiquitous performance issues in public WMSs. (3) The response time shows spatiotemporal patterns. Our experiment results indicate a positive correlation between WMS response time and the spatial distance from users to servers. The closest monitoring site tends to have the smallest average response time. Furthermore, the shorter the distance, the slighter the fluctuation and the more stable the response time. The trend in the response time series fluctuates significantly with local network traffic and synthesizes the minor random variances. These findings are important to understand the factors affecting service performance. Our investigation provides a valuable guideline for selecting WMS resources and optimizing WMS performance.

*5.2. Suggestion and Future Work*

To improve WMS discovery, selection and application, suggestions for standards developers and our potential future research directions include:

(1) Redesigning or redefining the WMS standard and improving both client-side and server-side functionality. The current WMS standard provides very simple and easy-to-use operations to retrieve maps rendered with widely-used industrial image formats on the server side. As web technologies develop, the computing and interactive capability of web browsers are becoming more powerful. Under such circumstances, the WMS standard should be refined, leaving more fine-grained control authority free on the client side for enhanced interactivity and visual analytics functions. For example, new operations can be added to provide access and interaction capability to manipulate individual features and layers. Rendering and animation can be customized on the client side, e.g., the style of map symbols. Meanwhile, the server side could improve performance, concurrent access capacity and validation functions for metadata.

(2) Building sophisticated WMS quality models. Response time prediction can facilitate service selection for time-critical applications. By analyzing the key impact factors and utilizing the spatiotemporal patterns of response times, prediction models could be built to achieve precise prediction. To support quality-driven WMS resource discovery, a comprehensive evaluation quality model could consider more quality metrics, e.g., data quality of maps, user feedback and preferences.

(3) Developing a state-of-the-art web portal for better service discovery. Interactive query and visual analytics functions must be enhanced for the next generation of geospatial web portals. Firstly, quality (e.g., performance) and user scoring should be integrated and supported as search criteria. Secondly, service comparisons and the visual analytics function should be enabled. For example, users could be permitted to compare the response time, user feedback and successability of selected services visually, in an interactive way.

(4) Optimizing the proposed monitoring framework. The scalability and flexibility of our distributed framework could be improved with a larger number of monitoring sites and services. More types of geospatial web services (e.g., ESRI RESTful services, OGC CSW, OGC Sensor Observation Service, OGC Web Processing Service) and operations should be supported.

**Acknowledgments:** This work is supported by the National Natural Science Foundation of China (No. 41501434 and No. 41371372), the Research Foundation of Wuhan University (No. 2042014kf0026) and the Natural Science Foundation of Hubei Province (No. 2015CFA053). Thanks to Mr. Steve McClure for language assistance.

**Author Contributions:** Zhipeng Gui designed the methods and wrote the paper. Jun Cao implemented the monitoring system and performed the experiments. Xiaojing Liu analyzed the data. Xiaoqiang Cheng and Huayi Wu contributed to the conceptualization and methods.

**Conflicts of Interest:** The authors declare no conflict of interest.